\documentclass[conference]{IEEEtran}
\IEEEoverridecommandlockouts
\usepackage{cite}
\usepackage{amsmath,amssymb,amsfonts}
\usepackage{algorithmic}
\usepackage{graphicx}
\usepackage{textcomp}
\usepackage{xcolor}
\usepackage{hyperref}
\usepackage{algorithm}
\usepackage{tabularx,booktabs}
\newcolumntype{Y}{>{\centering\arraybackslash}X}
\usepackage{multirow}
\usepackage[font=small,labelfont=bf]{caption}
\def\BibTeX{{\rm B\kern-.05em{\sc i\kern-.025em b}\kern-.08em
    T\kern-.1667em\lower.7ex\hbox{E}\kern-.125emX}}
\begin{document}

\title{How Potent are Evasion Attacks for Poisoning Federated Learning-Based Signal Classifiers? \\
\thanks{*S. Wang and R. Sahay contributed equally to this work.}
}

\author{Su Wang*, Rajeev Sahay*, and Christopher G. Brinton \\ Elmore Family School of Electrical and Computer Engineering, Purdue University \\ \{wang2506,sahayr,cgb\}@purdue.edu}

\maketitle

\begin{abstract}

There has been recent interest in leveraging federated learning (FL) for radio signal classification tasks. In FL, model parameters are periodically communicated from participating devices, training on their own local datasets, to a central server which aggregates them into a global model. While FL has privacy/security advantages due to raw data not leaving the devices, it is still susceptible to several adversarial attacks. In this work, we reveal the susceptibility of FL-based signal classifiers to model poisoning attacks, which compromise the training process despite not observing data transmissions. In this capacity, we develop an attack framework in which compromised FL devices perturb their local datasets using adversarial evasion attacks. As a result, the training process of the global model significantly degrades on in-distribution signals (i.e., signals received over channels with identical distributions at each edge device). We compare our work to previously proposed FL attacks and reveal that as few as one adversarial device operating with a low-powered perturbation under our attack framework can induce the potent model poisoning attack to the global classifier. Moreover, we find that more devices partaking in adversarial poisoning will proportionally degrade the classification performance. 

\end{abstract}


\section{Introduction}


As the Internet of Things (IoT) expands, efficient management of the wireless spectrum is critical for next-generation wireless networks. Intelligent signal classification (SC) techniques, such as automatic modulation classification (AMC), are a key technology for enabling such efficiency in the increasingly crowded radio spectrum. Such methods dynamically predict signal characteristics, such as its modulation scheme, direction of arrival, and channel state information (CSI), using the in-phase and quadrature (IQ) time samples of received signals. Deep learning is known to be highly effective for SC, outperforming likelihood-based classifiers without requiring specific feature engineering of the IQ samples \cite{dl_amc}. 

Federated learning (FL) \cite{fl}, a technique for distributing model training, has recently been considered for DL-based SC~\cite{amc_fl1}. 
In FL-based SC, each participating device trains a model on their locally collected dataset of received signals. Periodically, each local device transmits their model parameters to a global server, which aggregates all the received model parameters. The global server then communicates the updated aggregated model to all participating devices. The participating FL devices (i.e., clients) resume training from the received model parameters returned from the global server. As a result of this design, locally received/collected signals are never transmitted over the network, as required by centralized SC, thus mitigating the potential of data leakage. 

Although FL does not directly transmit client datasets, it is still susceptible to adversarial attacks at the data level. In this work, we reveal the degree of vulnerability existing FL-based SC tasks have to such attacks. Specifically, we develop an attack framework in which adversarial evasion perturbations \cite{evasion_atks} are used to conduct model poisoning attacks \cite{model_poison} in FL-based SC networks. In this capacity, we consider an FL SC network in which a subset of the participating clients intentionally perturb their local datasets in an effort to force the corresponding local model to learn a shifted distribution of the true received signals. We show that the training convergence and classification performance of the global model is significantly degraded as a result of the evasion perturbations induced on the subset of local clients. 


\textbf{Related Work:} Centralized DL-based SC has been shown to be susceptible to adversarial evasion attacks \cite{adv_in_rf,amc_adv_atk1,amc_adv_atk2,amc_adv_atk3}. In these settings, the SC DL classifier is attacked during the inference phase. Specifically, the classifiers are first trained using a collection of labeled radio signals. Then, during test time, the adversary perturbs inputs to induce the trained classifier to output erroneous predictions. Several defenses have been proposed to mitigate such attacks \cite{amc_def1,amc_def2}, but these methods are designed specifically for test-time attacks in the centralized SC scenario. Our focus, on the other hand, is on model poisoning (as opposed to data poisoning) attacks instantiated during the training phase of FL-based SC. 

One very effective technique for mitigating evasion attacks on centralized SC systems is adversarial training \cite{adv_trn1,adv_trn2}, where the training set is augmented with adversarial examples in order to increase test-time performance in the presence of such attacks. However, adversarial training on samples with high-bounded perturbations results in the model overfitting to adversarial examples, thus reducing classification performance on unperturbed samples \cite{adv_trn_overfitting}. In this work, we utilize this property by augmenting the local training set of particular FL devices with imperceptible adversarial evasion attacks. This, in turn, poisons the global model during training, thus reducing its classification performance. 

In the FL context, model poisoning attacks, which aim to corrupt the training process, have been proposed for image processing tasks \cite{atks_in_fl}. Such attacks consist of label flipping \cite{label_flip} and model parameter perturbations \cite{byz_atk}. In the former case, the resulting attack potency is low and can be mitigated through global averaging of all model parameters. The latter case relies on perturbing weights after training, which can be detected using existing distributed SC algorithms \cite{fl_atk_det}. Contrary to these works, our proposed attack framework does not rely on perturbing the model parameters after local training, thus bypassing detection mechanisms from previously proposed SC frameworks. To the best of our knowledge, this is the first work to propose such a model poisoning attack for FL-based SC.

\textbf{Summary of Contributions:} The main contributions of this work are as follows: 

\begin{enumerate}
    \item \textbf{Vulnerability of FL signal classifiers to adversarial poisoning:} We reveal the susceptibility of FL-based signal classifiers to model poisoning attacks. 
    
    \item \textbf{FL-Based SC Poisoning Framework:} We propose a framework for poisoning the training process of FL-based SC by perturbing local datasets using potent, but imperceptible, adversarial evasion attacks. 
    
    \item \textbf{Experimental Validation of FL-Based SC Poisoning:} We demonstrate the susceptibility of signal classifiers to model poisoning through numerical experiments with a real-world AMC dataset in different adversarial environments (e.g., networks of varying size or adversaries). 
    

\end{enumerate}

\begin{figure}[t]
    \centering
    \includegraphics[width=0.48\textwidth]{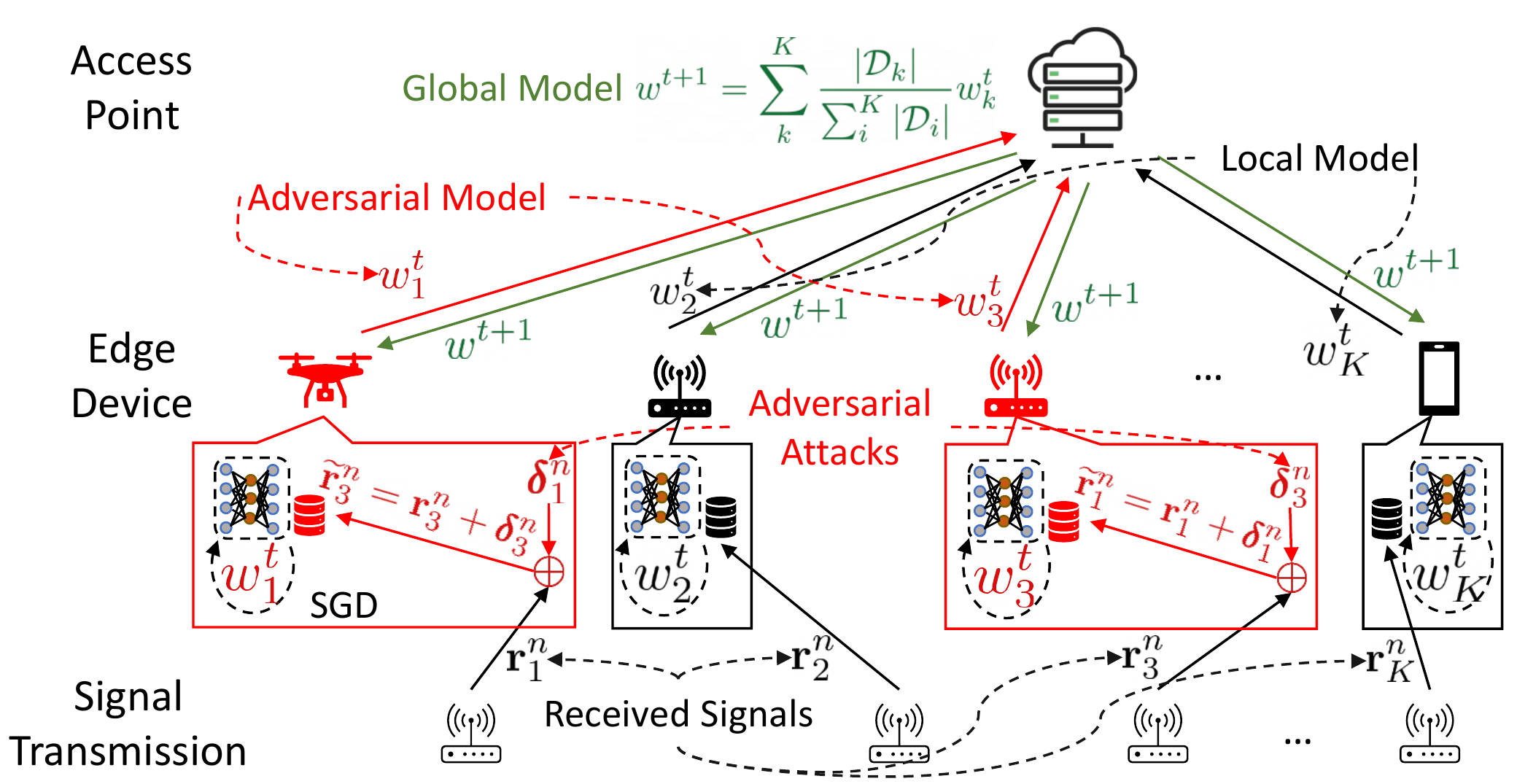}
    \caption{Our proposed FL-based SC framework in which select edge devices train their local models on datasets perturbed with adversarial evasion attacks. As a result, the global model jointly performs aggregation on poisoned and unpoisoned model parameters and subsequently distributes the now-poisoned ML model.} 
    \label{fig:sys_model}
    \vspace{-4mm}
\end{figure}

\vspace{-0.2cm}
\section{Methodology}

In this section, we begin by discussing the preliminary notations and system variables in Sec. \ref{sec:sig_mod}. Next, we describe the data perturbation process in Sec. \ref{sec:atk_fw} and, finally, we give the overall FL-based model poisoning framework for SC in Sec. \ref{sec:mod_pois}. Our proposed framework described throughout this section is shown in Fig. \ref{fig:sys_model}. 

\subsection{Signal and Classifier Modeling} \label{sec:sig_mod}

We consider an FL framework consisting of $k = 1, 2, \ldots, K$ participating training devices, where each device contains a local dataset denoted by $\mathcal{D}_{k}$ consisting of $|\mathcal{D}_{k}|$ samples. At each device, $\mathcal{D}_{k}$ is comprised of a set of received signals, which were each transmitted to device $k$ through the channel $\mathbf{h}_{k} = [h_{k}[0],\ldots,h_{k}[\ell - 1]]^{T}$, where $\ell$ is the length of the received signal's observation window. We assume that the channel distribution between the transmitter and each device is independent and identically distributed (i.i.d.). Formally, the $n^{\text{th}}$ signal received at device $k$ is modeled by
\begin{equation}
    \mathbf{r}^{n}_{k} = \sqrt{\rho}\mathbf{H}_{k}\mathbf{s} + \mathbf{n},
\end{equation}
where $\mathbf{s} = [s[0], \ldots, s[\ell - 1]]$ is the transmitted signal, $\mathbf{H}_{k} = \text{diag}\{h_{k}[0],\ldots,h_{k}[\ell - 1]\} \in \mathbb{C}^{\ell \times \ell}$, $\mathbf{n} \in \mathbb{C}^{\ell}$ is complex additive white Gaussian noise (AWGN), and $\rho$ denotes the signal to noise ratio (SNR), which is known at the receiver of each device. Each realization of $ \mathbf{r}^{n}_{k}$ of various constellations, and the FL objective is to learn a global signal classifier by training all local models to classify the signal as one of $C$ possible signal constellations. 

Although all received signals are complex, $\mathbf{r}^{n}_{k} \in \mathbb{C}^{\ell}$, we denote each signal in terms of its real and imaginary components, $\mathbf{r}^{n}_{k} \in \mathbb{R}^{\ell \times 2}$, where the two columns represent the real and imaginary components of $\mathbf{r}^{n}_{k}$, in order to (i) utilize all signal features during training and (ii) use real-valued DL architectures as overwhelmingly used in DL and FL-based SC. 

At the beginning of each training round, $t$, the global model transmits its parameters, $w^{t}$, to each FL device. Each device then trains, using $w^{t}$ as the starting point, its own local model, denoted by $f(\cdot, w^{t}): \mathbb{R}^{\ell \times 2} \rightarrow \mathbb{R}^{C}$, where $f()$ denotes the deep learning classifier (identical architecture at each device) and $(\cdot)$ represents the input. At the termination of the training round, each local device returns $w^{t}_{k}$, which is the model parameters of device $k$ after the completion of training round $t$ on $\mathcal{D}_{k}$, to the global server for aggregation (further discussed in Sec. \ref{sec:mod_pois}). After aggregation, the global server transmits the updated model parameters, $w^{t+1}$, for the next round of training. The model prediction, after training round $t$, is given by $\hat{\mathbf{y}}^{n}_{k} = f(\mathbf{r}^{n}_{k}, w^{t})$, where $\hat{\mathbf{y}}^{n}_{k} \in \mathbb{R}^{C}$ denotes the predicted output vector of $\mathbf{r}^{n}_{k}$ from $f()$. Moreover, the predicted signal constellation is given by $\text{argmax}_{j} \hspace{1mm} \hat{\mathbf{y}}^{n}_{k, j}$, where $\hat{\mathbf{y}}^{n}_{k, j} \in \mathbb{R}$ is the $j^{\text{th}}$ element of $\hat{\mathbf{y}}^{n}_{k}$. 

\subsection{Local Data Perturbation Generation} \label{sec:atk_fw}

Here, we describe the process followed by each adversarial device (we show the effects of varying the number of adversarial devices that follow this procedure in Sec. \ref{sec:perf_eval}). Within our proposed framework, a subset of devices, termed \emph{adversarial devices}, will train on perturbed inputs, beginning at training iteration $t_0$, and the remaining devices will continue training on their original unperturbed local datasets. At the beginning of each training iteration, after the local model has received an updated global model, adversarial devices will craft adversarial evasion perturbations on each instance of $\mathbf{r}^{n}_{k}$. The $n^{\text{th}}$ resulting sample is denoted by 
\begin{equation} \label{adv_sig}
    \widetilde{\mathbf{r}}^{n}_{k} = \mathbf{r}^{n}_{k} + \pmb{\delta}^{n}_{k},
\end{equation}
where $\pmb{\delta}^{n}_{k}$ is the adversarial perturbation crafted for the $n^{\text{th}}$ signal on device $k$. 

The adversarial perturbation, $\pmb{\delta}^{n}_{k}$, could be crafted at each local device by utilizing common perturbation models such as AWGN or changing the local data completely by e.g., using zero-vectors as training samples or training on signals received from an out-of-distribution channel.
However, the injection of AWGN results in less potent attacks to the global model in comparison to our proposed perturbation methodology (as we will show in Sec. \ref{sec:perf_eval}). On the other hand, although changing the local training data may result in more potent attacks, the global model can simply query training samples from each local device to identify the adversarial device. Using adversarial evasion attacks, as we propose, induces a higher attack potency while simultaneously being imperceptible and, thus, is able to withstand FL adversarial attack detectors that rely on querying local data to identify adversarial devices. 


To craft an effective and imperceptible perturbation, each adversarial device will aim to satisfy
\begin{subequations} \label{adv_opt:all-lines}
\begin{align}
    \underset{\pmb{\delta}}{\text{min}} \quad &  ||\pmb{\delta}^{n}_{k}||_{2} \label{adv_opt:line_1} \\
    \text{s. t.} \quad &  \hspace{0.5mm} f(\mathbf{r}^{n}_{k}, w^{t}) \neq f(\mathbf{r}^{n}_{k} + \pmb{\delta}^{n}_{k}, w^{t}), \label{adv_opt:line_2} \\ 
    \quad & ||\pmb{\delta}^{n}_{k}||_{2}^{2} \leq P, \label{adv_opt:line_3} \\
     \quad & \mathbf{r}^{n}_{k} + \pmb{\delta}^{n}_{k} \in \mathbb{R}^{\ell \times 2} \label{adv_opt:line4}, 
\end{align}
\end{subequations} 
where $||\cdot||_{2}$ denotes the $l_{2}$ norm and $P$ is the power budget for each adversarial signal. In (\ref{adv_opt:all-lines}), (\ref{adv_opt:line_1}) minimizes the needed power in order to keep the perturbation imperceptible, (\ref{adv_opt:line_2}) aims to change the prediction of the perturbed sample from the weights received at the beginning of training, (\ref{adv_opt:line_3}) restricts the adversary's power budget, and (\ref{adv_opt:line4}) ensures that $\widetilde{\mathbf{r}}^{n}_{k}$ remains in the same dimensional space as $\mathbf{r}^{n}_{k}$. 

\begin{algorithm}[h!] 
   \caption{FL model poisoning in SC at training iteration $t > t_{0}$, where the adversarial devices train on perturbed inputs.}
   \label{adv_alg}
   \begin{algorithmic}[1] 
        \STATE \textbf{input:} $w^{t}$: Global parameter values at training round $t$ \\ 
        

            \FOR{$q = 1, \ldots, Q$} 
                \STATE $\mathcal{B} \gets$ split $\mathcal{D}_{q}$ into batches
                
                \FOR{batch $b \in \mathcal{B}$}
                    \STATE $w^{t}_{q} \gets w^{t} - \eta_{k} \hspace{1mm} \nabla_{w^{t}}\mathcal{L}(b, w^{t})$
                    
                \ENDFOR
            \ENDFOR

            \FOR{$a = 1, \ldots, A$}
                \STATE $\widetilde{\mathcal{D}}_{a} \gets \{\}$
                \FOR{$\mathbf{r}^{n}_{a} \in \mathcal{D}_{a}$}
                    \STATE $\pmb{\delta}^{n}_{a} = \sqrt{P} \frac { \nabla_{\mathbf{r}_{\text{k}}^{n}} \mathcal{L}(\mathbf{r}^{n}_{k}, \mathbf{y}^{n}_{k}, w^{t})} {|| \nabla_{\mathbf{r}_{\text{k}}^{n}} \mathcal{L}(\mathbf{r}^{n}_{k}, \mathbf{y}^{n}_{k}, w^{t})||_{2}}$
                    \STATE $\widetilde{\mathbf{r}}^{n}_{a} = \mathbf{r}^{n}_{a} + \pmb{\delta}^{n}_{a}$
                    \STATE add $\widetilde{\mathbf{r}}^{n}_{a}$ to $\widetilde{\mathcal{D}}_{a}$
                    
                \ENDFOR
                
                \STATE $\widetilde{\mathcal{B}} \gets$ split $\widetilde{\mathcal{D}}_{a}$ into batches
                
                \FOR{batch $\widetilde{b} \in \widetilde{\mathcal{B}}$}
                    \STATE $\widetilde{w}^{t}_{a} \gets w^{t} - \eta_{k} \hspace{1mm} \nabla_{w^{t}}\mathcal{L}(\widetilde{b}, \widetilde{w}^{t})$
                    
                \ENDFOR

            \ENDFOR

            \STATE $w^{t+1} = \sum_{a}^{A} \frac{|\mathcal{D}_{a}|}{\sum_{i}^{K} |\mathcal{D}_{i}|} \alpha_{a} \widetilde{w}^{t}_{a} + \sum_{q}^{Q} \frac{|\mathcal{D}_{q}|}{\sum_{i}^{K} |\mathcal{D}_{i}|} w^{t}_{q}$

        \RETURN $w^{t+1}$
  
  \end{algorithmic}
\end{algorithm}

Due to its excessive nonlinearity, however, (\ref{adv_opt:all-lines}) is difficult to solve using traditional optimization methods. Thus, we approximate its solution using the fast gradient sign method (FGSM) \cite{fgsm}. The FGSM perturbation for our proposed FL-based SC model is given by
\begin{equation} \label{fgsm}
    \pmb{\delta}^{n}_{k} = \sqrt{P} \frac { \nabla_{\mathbf{r}^{n}_{\text{k}}} \mathcal{L}(\mathbf{r}^{n}_{k}, \mathbf{y}^{n}_{k}, w^{t})} {|| \nabla_{\mathbf{r}^{n}_{\text{k}}} \mathcal{L}(\mathbf{r}^{n}_{k}, \mathbf{y}^{n}_{k}, w^{t})||_{2}},
\end{equation}
where 
\begin{equation}
    \mathcal{L}(\mathbf{r}^{n}_{k}, \mathbf{y}^{n}_{k}, w^{t}) = \sum_{j=1}^{C} \mathbf{y}^{n}_{k, j} \text{log}(\hat{\mathbf{y}}^{n}_{k})
\end{equation}
is the cross entropy loss with $\mathbf{y}^{n}_{k, j}$ denoting the $j^{\text{th}}$ element of the true label vector corresponding to $n^{\text{th}}$ sample on the $k^{\text{th}}$ device and $\nabla_{\mathbf{r}_{\text{k}}} \mathcal{L}(\mathbf{r}^{n}_{k}, \mathbf{y}^{n}_{k}, w^{t})$ denotes the gradient of $\mathcal{L}(\mathbf{r}^{n}_{k}, \mathbf{y}^{n}_{k}, w^{t})$ w.r.t. $\mathbf{r}^{n}_{k}$. Finally, $\sqrt{P} / || \nabla_{\mathbf{r}_{\text{k}}} \mathcal{L}(\mathbf{r}^{n}_{k}, \mathbf{y}^{n}_{k}, w^{t})||_{2}$ is the scaling factor used to satisfy the power constraint in (\ref{adv_opt:line_3}).

The objective of each adversarial device is to overfit their local model to the perturbed dataset generated using (\ref{adv_sig}) and (\ref{fgsm}) for each training sample. We will denote the batch of $N$ perturbed samples as $\widetilde{b} = \{(\widetilde{\mathbf{r}}^{n}_{k}, \mathbf{y}^{n}_{k})\}_{n=1}^{N}$ and the weights at the end of training round, $t$, on an adversarial device, $k$, as $\widetilde{w}^{t}_{k}$. Similarly the batch of $N$ unperturbed inputs as well as the weights at the end of training round $t$ at a non-adversarial device will be denoted as $b = \{(\mathbf{r}^{n}_{k}, \mathbf{y}^{n}_{k})\}_{n=1}^{N}$ and $w^{t}_{k}$, respectively. 


\subsection{Model Poisoning in FL-Based Signal Classification} \label{sec:mod_pois}

To begin each training round, $t$, in the FL AMC training process, the global model will transmit $w^{t}$ to each participating FL device. Note when $t=1$ (i.e., the first round of training), $w^{1}$ is randomly initialized. After receiving $w^{t}$, each FL device will train $f(\cdot, w^{t})$ on $\mathcal{D}_{k}$. The model parameters of the $a^{\text{th}}$ adversarial device will be updated, beginning on training round $t_{0}$, according to 
\begin{equation}
    \widetilde{w}^{t}_{a} = w^{t} - \eta_{k} \hspace{1mm} \nabla_{w^{t}}\mathcal{L}(\widetilde{b}, \widetilde{w}^{t}), 
\end{equation}
while the model parameters of the $q^{\text{th}}$ non-adversarial device, along with adversarial devices prior to training round $t_{0}$, will be updated according to 
\begin{equation}
    w^{t}_{q} = w^{t} - \eta_{k} \hspace{1mm} \nabla_{w^{t}}\mathcal{L}(b, w^{t}), 
\end{equation}
where $\eta_{k}$ is the learning rate at device $k$. At the termination of training round $t$, each FL device will transmit its updated model parameters back to the global server. Although non-adversarial devices will transmit $w^{t}_{q}$ to the global model, adversarial devices will transmit $\alpha_{a} \hspace{0.5mm} \widetilde{w}^{t}_{a}$, where $\alpha_{a} > 0$ is a scaling factor used at adversarial device $a$ that can be used to make the effect of the perturbed weights more potent at the global model. Note that $\alpha_{a} = 1$ corresponds to not scaling the trained weights. In addition, each FL device will also transmit $|\mathcal{D}_{k}|$ to the global model for appropriate parameter scaling from each participating device during global aggregation. 

The global model will then perform global aggregation using the received weights. From the perspective of the global model, the aggregation scheme used to generate the model parameters to send to the local models for the next training iteration will be
\begin{equation}
    w^{t+1} = \sum_{k}^{K} \frac{|\mathcal{D}_{k}|}{\sum_{i}^{K} |\mathcal{D}_{i}|} w^{t}_{k},
\end{equation}
where $K$ is the total number of FL devices. However, the true aggregation process, taking the effect of the adversarial devices into account, is given by
\begin{equation} \label{true_avg}
    w^{t+1} = \sum_{a}^{A} \frac{|\mathcal{D}_{a}|}{\sum_{i}^{K} |\mathcal{D}_{i}|} \alpha_{a} \widetilde{w}^{t}_{a} + \sum_{q}^{Q} \frac{|\mathcal{D}_{q}|}{\sum_{i}^{K} |\mathcal{D}_{i}|} w^{t}_{q}, 
\end{equation}
where $A$ and $Q$ are the total number of adversarial and non-adversarial devices, respectively, and $K = A + Q$. As a result of this design, the global model will suffer in convergence performance despite not aggregating the local data to a centralized location. The complete overview of our model poisoning framework is given in Algorithm \ref{adv_alg}.


\begin{figure}[t]
    \centering
    \includegraphics[width=0.48\textwidth]{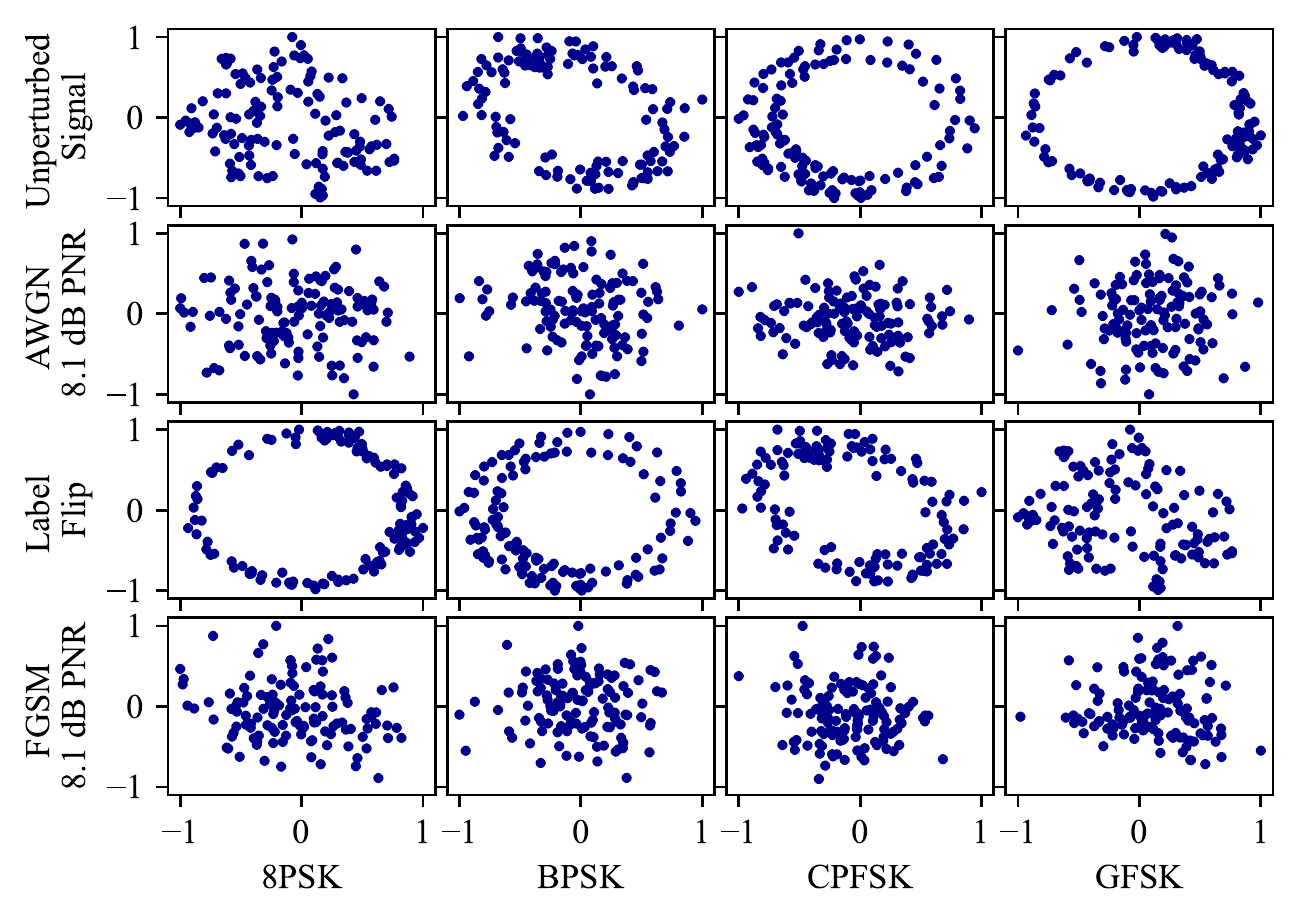}
    \caption{Waveform constellation visualizations for four RML labels: 8PSK, BPSK, CPFSK, and GFSK. The top row depicts received signals with no additive perturbations added at the local device. The AWGN and FGSM perturbations are both shown with $\text{PNR}=8.1$ dB and visually appear similar. The label flipping attack, shown in the third row, changes the true underlying label, for example the BPSK and CPFSK constellations are flipped.}
    \label{fig:data_const}
    \vspace{-2mm}
\end{figure}

\section{Results and Discussion}

Here, we begin in Sec. \ref{sec:dataset} by discussing the DL classification architecture employed at each local model as well as the dataset used in our evaluation. Then, in Sec. \ref{sec:perf_eval}, we present the results of our numerical simulations in which we consider a variety of different adversarial operating environments. 

\subsection{FL Classification Architecture and Dataset} \label{sec:dataset}
Each device trains a local DL classifier using a lightweight version (for computational efficiency) of the VT-CNN2 architecture \cite{dl_amc}. Specifically, each local classifier is composed of 2 sequential convolutional layers with 16 and 80 feature maps, consisting of $1 \times 3$ and $2 \times 3$ kernel sizes, respectively, followed by a 256 unit dense layer and a $C$ dimensional output layer. Each intermediate layer applies the ReLU activation, and the output layer applies the softmax activation. Thus $\hat{\mathbf{y}}^{n}_{k, j}$ can be interpreted as the probability of the $n^{\text{th}}$ input from the $k^{\text{th}}$ device belonging to the $j^{\text{th}}$ class. 
We use $\eta_k = 0.001 \hspace{1mm} \forall \hspace{1mm} k$, and we set $\alpha_a = 1 \hspace{1mm} \forall \hspace{1mm} a$ to isolate the effect of evasion attacks.

To evaluate our poisioning framework, we employ the RadioML2016.10a dataset (RML), which is an automatic modulation classification (AMC) dataset commonly used to benchmark the effectiveness of wireless communications algorithms for radio signal classification. The dataset consists of signals in the following ten modulation constellations stored at an SNR of 10 dB: 8PSK, AM-DSB, BPSK, CPFSK, GFSK, PAM4, QAM16, QAM64, QPSK, and WBFM. In total, we apply a $75\%/25\%$ train/test split, resulting in $45$K training samples, split among the participating clients, and $15$K testing samples contained at the global server. 


Each RML signal is normalized to unit energy and has observation window of length $\ell = 128$. We depict the RML constellations in the uppermost row of Fig.~\ref{fig:data_const}, and show the signals after perturbing using FGSM as well as after perturbing using AWGN (baseline) and label flipping (baseline) in Fig.~\ref{fig:data_const}. 

We measure the potency of the local perturbations in terms of the perturbation to noise ratio (PNR) given by
\begin{equation} \label{eq:pnr}
    \text{PNR [dB]} = \text{PSR [dB]} + \text{SNR [dB]},
\end{equation}
where PSR is the perturbation to signal ratio. 

\begin{figure*}[t]
    \centering
    \includegraphics[width=0.99\textwidth]{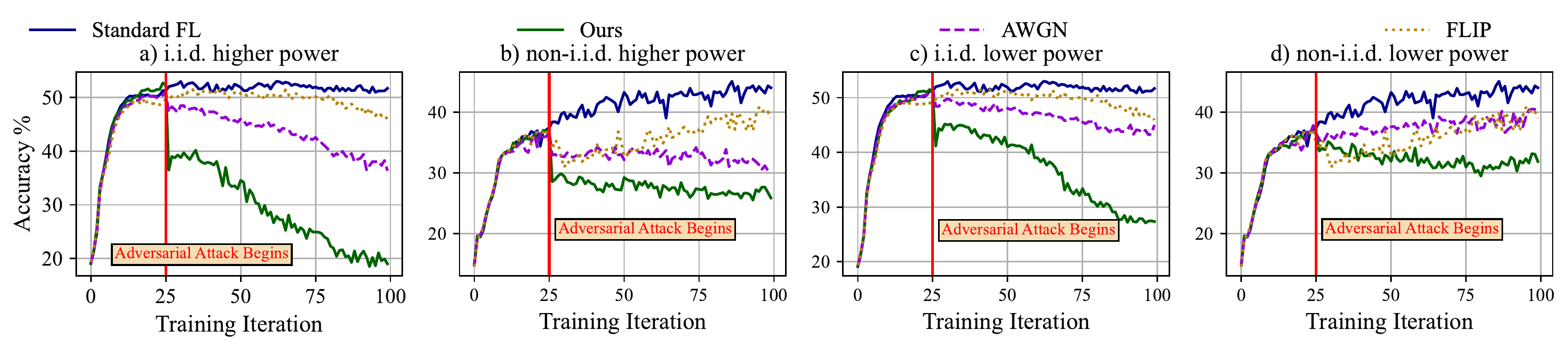}
    \caption{Training performance of high and low power perturbations for a network with $30\%$ adversarial devices. High power perturbations at $8.1$ dB PNR are shown in a) and b) and low power perturbations at $6.7$ dB PNR are shown in c) and d). Lower accuracy indicates higher adversarial impact. For both i.i.d. and non-i.i.d. scenarios, our algorithm yields the most potent model poisoning attack.} 
    \label{fig:all_ovr}
    \vspace{-1mm}
\end{figure*}


\begin{figure*}[t]
    \centering
    \includegraphics[width=0.99\textwidth]{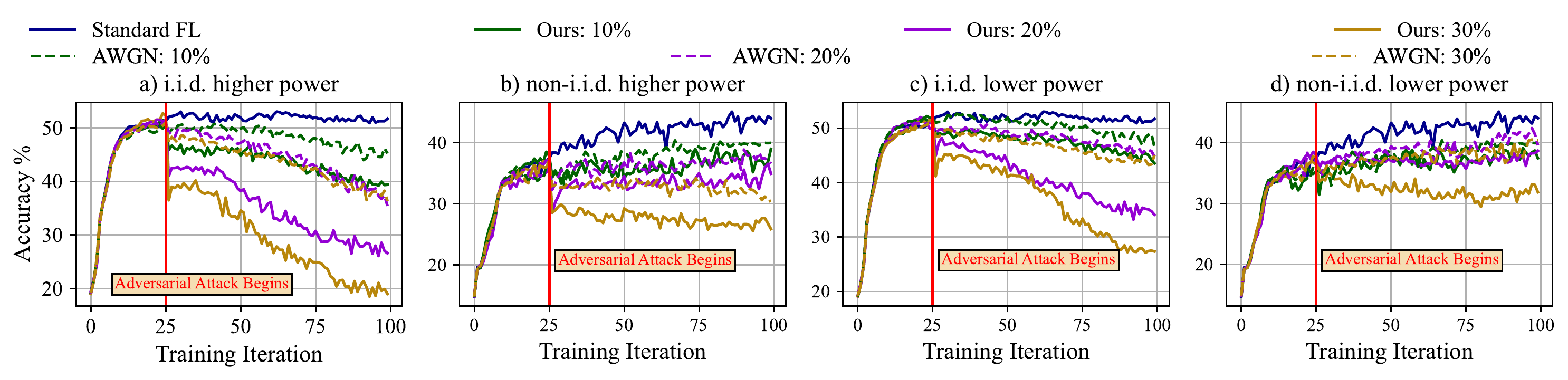}
    \caption{Varying the proportion of network compromised by adversaries from $10\%$ to $30\%$.  
    High power perturbation ($8.1$ dB PNR) experiments are in a) and b) and low power perturbations ($6.7$ dB PNR) experiments are in c) and d). 
    Our algorithm remains the most potent even as the quantity of adversarial devices varies, indicated within the legend. }
    \label{fig:all_scaled}
    \vspace{-1mm}
\end{figure*}

\subsection{Performance Evaluation} \label{sec:perf_eval}
We first demonstrate the effectiveness of our method, relative to the baselines, in Sec~\ref{sec:baseline_all}. 
We then show that evasion-based attacks directly scale with the number of adversaries in the network in Sec~\ref{sec:num_advs}. 

Unless otherwise stated, in our evaluations, we consider a network of $K=10$ devices consisting of classifiers based on the lightweight VT-CNN2 architecture described in Sec. \ref{sec:dataset} in both i.i.d. and non-i.i.d. signal distributions among devices. In an i.i.d. environment, our network devices all contain the same quantity of local data and have local data sampled uniformly at random from each class of the full training dataset.
In the non-i.i.d. case, devices have data quantity chosen randomly from $\mathcal{N}(4500,45)$ and data randomly sampled from only three labels as in~\cite{wang2021device}. After training iteration $t_{0} = 25$, $30\%$ of the network is compromised by adversarial influence, and begins training on perturbed local datasets.

\subsubsection{Baseline Comparison} \label{sec:baseline_all}
We compare our framework to two baseline methods: data poisoning via AWGN and label flipping (FLIP). AWGN attacks inject random Gaussian noise into the training data at the local devices while label flipping intentionally mislabels local training data. 
We choose to compare against these baselines since, similar to our method, they both rely on intentional manipulations of local training data to poison model aggregations and thus the global model. 


In our evaluation, we vary the power of the perturbation to assess its effect on the potency of the attack. For the high power scenario in Fig.~\ref{fig:all_ovr}a) and~\ref{fig:all_ovr}b), we set the PNR of both our method as well as the AWGN baseline to $8.1$ dB. Our methodology yields the most potent training performance for both i.i.d. and non-i.i.d. scenarios. In the i.i.d. case in Fig.~\ref{fig:all_ovr}a), our algorithm reduces classification performance, at the termination of training, by over $31\%$, which is a $17\%$ increase over AWGN and a $26\%$ increase over FLIP. Similarly, in the non-i.i.d. case in Fig.~\ref{fig:all_ovr}b), our method reduces the accuracy by over $17\%$, which is $4\%$ more than AWGN and $13\%$ more than FLIP. 

In the low power scenario in Fig.~\ref{fig:all_ovr}c) and~\ref{fig:all_ovr}d), the AWGN attack and our algorithm both have $6.7$ dB PNR. 
FLIP is PNR independent and, thus, has the same results from Fig.~\ref{fig:all_ovr}a) and~\ref{fig:all_ovr}b). 
Here, our methodology continues to outperform all baselines. 
In the i.i.d. case, we reduce the classification performance, at the termination of training, by over $23\%$, which corresponds to a $15\%$ improvement over AWGN and a $16\%$ improvement over FLIP. Furthermore, in the non-i.i.d. scenario, our algorithm reduces performance by $12\%$, outperforming, potency-wise, AWGN by $8\%$ and FLIP by $8\%$. 

The reduction in nominal impact of all evasion attacks in non-i.i.d. environments seen throughout Fig.~\ref{fig:all_ovr} is the result of an innate property of FL. 
Specifically, in non-i.i.d. scenarios, devices and thus adversaries may not have data from all possible labels. As a result, the adversaries can only bias the ML model's classification performance on the specific labels that they have corresponding data for. 
Consequently, after model aggregations, the global ML model display weaker classification on underlying labels present at the adversaries. 



\subsubsection{Network Scaling Effects} \label{sec:num_advs}
Next, we evaluate the potency of our framework in Fig.~\ref{fig:all_scaled}, where we vary the proportion of the network compromised by adversarial devices from $10\%$ to $30\%$. 
Label flipping attacks do not necessarily become more potent as more adversaries enter the network, as a simple label flipping attack always attacks/flips the same labels and, as a result, the total number of compromised labels is independent of the number of adversaries. 
Therefore, to clearly demonstrate the impact of more compromised devices, we omit the FLIP attack. 
Otherwise, the experimental setup for Fig.~\ref{fig:all_scaled} remains the same as that for Fig.~\ref{fig:all_ovr}.

In the high power scenario in Fig.~\ref{fig:all_scaled}a) and~\ref{fig:all_scaled}b), our method reduces final classification performance by $11\%$ to over $31\%$ as the proportion of adversarial devices increases from $10\%$ to $30\%$ for the i.i.d. case, and by $5\%$ to over $17\%$ as the proportion of adversarial devices increases from $10\%$ to $30\%$ for the non-i.i.d. case. 
In the low power case of Fig.~\ref{fig:all_scaled}c) and~\ref{fig:all_scaled}d), our algorithm reduces final accuracy by $8\%$ to over $23\%$ as the proportion of adversarial devices increases from $10\%$ to $30\%$ for the i.i.d. scenario, and by $4\%$ to over $12\%$ as the proportion of adversarial devices increases from $10\%$ to $30\%$ for the non-i.i.d. scenario. 
In these cases, when the proportion of adversaries increases 3x, the degradation increases 3x. The classification degradation grows linearly with the proportion of adversaries. 
Regardless of the total number of network adversaries, our methodology continues to either outperform or match their baseline counterparts while remaining imperceptible to the global model. 

\begin{figure}[t]
    \centering
    \includegraphics[width=0.48\textwidth]{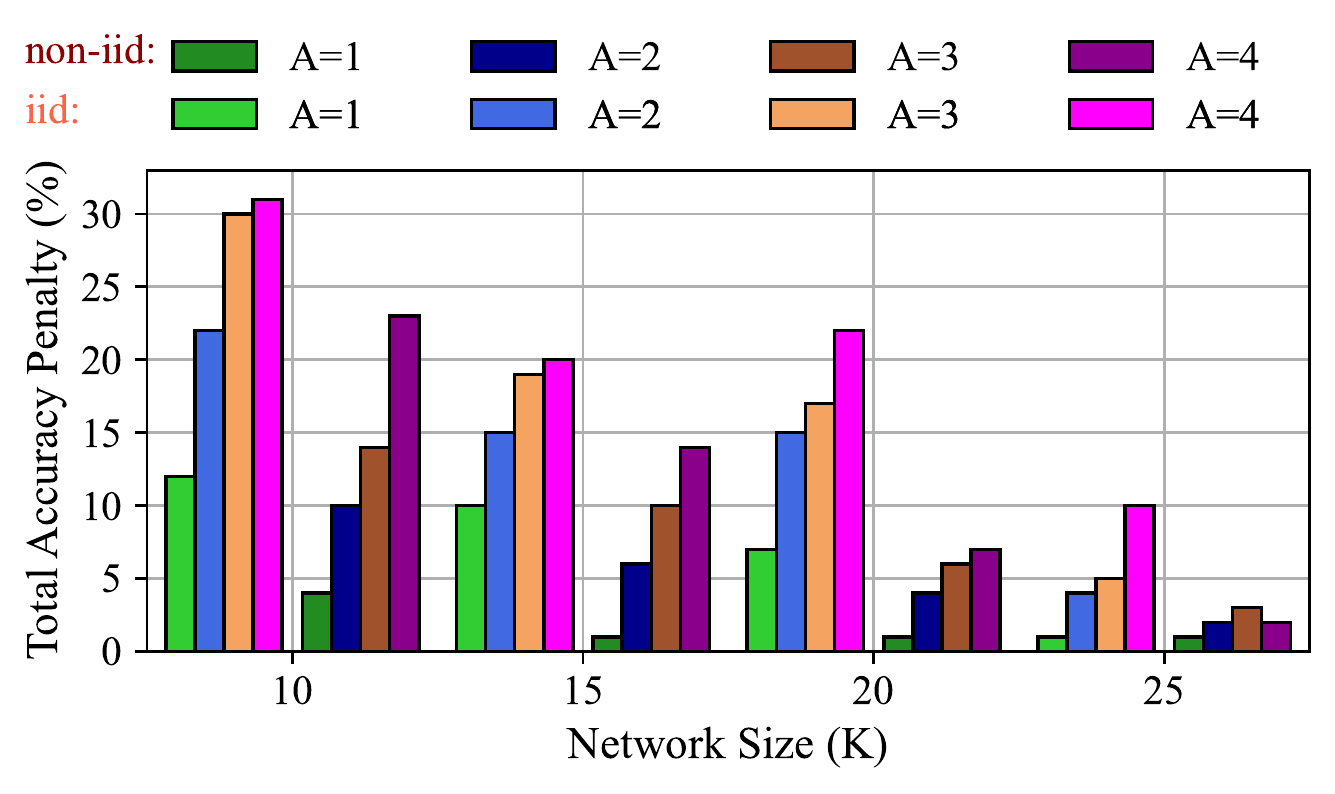}
    \caption{The impact of network size on reducing the classification performance. Light and dark colors indicate i.i.d. and non-i.i.d. underlying data distributions respectively. 
    Larger networks require more nominal adversaries to achieve significant accuracy penalties.}
    \label{fig:network_scale}
    \vspace{-3mm}
\end{figure}

We also investigate the impact of incrementing the quantity of adversarial devices, $A$, in networks of varying size. In Fig.~\ref{fig:network_scale}, to better capture the effect of adversaries in varying network sizes, we show the total accuracy penalty relative to the unperturbed scenario for four different network sizes. Increasing network size with a constant number of adversaries decreases the proportion of devices that are adversarial. 
Thus, to get the same level of adversarial influence in larger networks, we would nominally require more adversaries. 
Our results in Fig.~\ref{fig:network_scale} confirm this intuition for both the i.i.d. and non-i.i.d. cases, and solidifies the insights from Fig.~\ref{fig:all_scaled}. For example, $4$ adversaries in the i.i.d. case leads to over $30\%$ accuracy penalty in a network of $10$ devices leads but only $10\%$ accuracy penalty in a network of $25$ devices. A similar trend holds for the non-i.i.d. case with $4$ adversaries yielding roughly $22\%$ accuracy penalty in a network of $10$ devices but only $2\%$ in a network of $25$ devices. The nominal decrease in accuracy penalty in non-i.i.d. settings as compared to i.i.d. settings shown in Fig.~\ref{fig:network_scale} further confirms that non-i.i.d. scenarios are also more resilient to adversarial evasion attacks. The reasons for this are the same as those presented in Sec~\ref{sec:baseline_all}. Essentially, in the non-i.i.d. scenario, adversaries only contain data from a select subset of labels and, as a result, are only able to perturb the model's classification power on those specific labels. 

\section{Conclusion and Future Work}
The growing adoption of FL based methodologies to improve wireless signal classification has many potential benefits. However, there are specific challenges within wireless environments that can impede the performance and training of such methodologies. 
In this work, we showed that evasion attacks have the potential to poison FL-based signal classifiers. 
Specifically, we showed that evasion attacks are effective against FL, where compromising even a single device can damage the rest of the network and this, in effect, increases steadily with the number of adversaries. 
Such evasion attacks are also difficult to detect and defend against in wireless settings as they bear statistical and visual similarities to  additive white Gaussian noise, a common occurrence in wireless networks.

In future work, we plan on further characterizing the characteristics of large-scale FL networks in the presence of adversarial influence. Then, we will develop an efficient and effective defense mechanism for FL against such adversarial influence. 

\bibliography{references}

\bibliographystyle{IEEEtran}

\end{document}